# Dark Matter Halos of Luminous Active Galactic Nuclei from Galaxy–Galaxy Lensing with the HSC Subaru Strategic Program

Wentao Luo[1,2], John D. Silverman[1,3], Surhud More[1,4], Andy Goulding[5], Hironao Miyatake[1,6,7], Takahiro Nishimichi[1,8], Chiaki Hikage[1], Lalitwadee Kawinwanichakij[1], Junyao Li[1,2], Qinxun Li[9], Xiangchong Li[1], Elinor Medezinski[5], Masamune Oguri[1,10,11], Taira Oogi[1], and Cristobal Sifon[12]

[1] Institute for the Physics and Mathematics of the Universe (Kavli IPMU, WPI), UTIAS, Tokyo Institutes for Advanced Study, University of Tokyo, Chiba 277-8583, Japan; wtluo@ustc.edu.cn
[2] Department of Astronomy, School of Physical Sciences, University of Science and Technology of China, Hefei, Anhui 230026, People's Republic of China
[3] Department of Astronomy, School of Science, The University of Tokyo, 7-3-1 Hongo, Bunkyo, Tokyo 113-0033, Japan
[4] IUCAA, Post Bag 4 Ganeshkhind, Savitribai Phule Pune University Campus, Pune 411 007, India
[5] Department of Astrophysical Sciences, Peyton Hall Princeton University, 08544 NJ, USA
[6] Institute for Advanced Research, Nagoya University, Nagoya 464-8601, Japan
[7] Division of Physics and Astrophysical Science, Graduate School of Science, Nagoya University, Nagoya 464-8602, Japan
[8] Center for Gravitational Physics, Yukawa Institute for Theoretical Physics, Kyoto University, Kyoto 606-8502, Japan
[9] Department of Physics and Astronomy, University of Utah, INSCC, Salt Lake City, UT 84102, USA
[10] Research Center for the Early Universe, University of Tokyo, Tokyo 113-0033, Japan
[11] Department of Physics, University of Tokyo, Tokyo 113-0033, Japan
[12] Physics Institute, Pontificia Universidad Católica de Valparaíso, Valparaíso 2340025, Chile
*Received 2021 November 9; revised 2024 September 29; accepted 2024 October 10; published 2024 December 6*

## Abstract

We assess the dark matter halo masses of luminous active galactic nuclei (AGNs) over the redshift range 0.2–1.2 using galaxy–galaxy lensing based on imaging data from the Hyper Suprime-Cam Subaru Strategic Program (HSC-SSP). We measure the weak lensing signal of a sample of 48,907 AGNs constructed using HSC and Wide-field Infrared Survey Explorer photometry. As expected, we find that the lensing mass profile of total AGN sample is consistent with that of massive galaxies ($\log(M_*/h^{-2}M_\odot) \sim 10.61$). Surprisingly, the lensing signal remains unchanged when the AGN sample is split into four host galaxy stellar mass bins. Specifically, we find that the excess surface density of AGNs residing in galaxies with high stellar masses significantly differs from that of the control sample. We further fit a halo occupation distribution model to the data to infer the posterior distribution of parameters including the average halo mass. We find that the characteristic halo mass of the full AGN population lies near the knee ($\log(M_{\rm h}/h^{-1}M_\odot) = 12.0$) of the stellar-to-halo mass relation (SHMR). Illustrative of the results given above, the halo masses of AGNs residing in host galaxies with high stellar masses (i.e., above the knee of the SHMR) fall below the calibrated SHMR while the halo masses of the low stellar mass sample are more consistent with the established SHMR. These results indicate that massive halos with a higher clustering bias tend to suppress AGN activity, probably due to the lack of available gas.

## 1. Introduction

Supermassive black holes (SMBHs) power the most energetic phenomena in our Universe, i.e., active galactic nuclei (AGNs) and luminous quasars. Their formation and growth is intimately related to the galaxies in which they reside. This is seen in the observed relations of the masses of SMBHs to the properties (stellar mass, bulge mass, and velocity dispersion) of their host galaxies (see J. Kormendy & L. C. Ho 2013 for a review on this topic). Simply, the more massive black holes sit in the centers of galaxies with high stellar masses. Closely related, there is a preference for AGNs to be associated with massive galaxies that are still forming new stars (e.g., J. D. Silverman et al. 2009; J. R. Mullaney et al. 2012; Y. Xie et al. 2021); this may indicate a scenario where concurrent fueling of SMBHs and star formation is occurring through shared gas reservoirs (galactic scale or larger; A. Schulze et al. 2019; R. Carraro et al. 2020; J. Shangguan et al. 2020) and/or galaxy mergers (e.g., P. F. Hopkins et al. 2008; A. D. Goulding et al. 2018). Subsequently, the effective shutdown in mass growth of the SMBH and its host is attributed to some form of feedback from the SMBH itself (e.g., A. King & K. Pounds 2015; W. Ishibashi et al. 2021). With much attention on the low-redshift Universe, many efforts have begun to establish these relations out to the peak of the cosmic star formation history and beyond to the epoch of reionization.

There is also much interest to determine whether the growth of SMBHs may be influenced by their larger-scale environment (e.g., G. Kauffmann et al. 2004; E. T. Khabiboulline et al. 2014). In particular, clustering measurements of quasars have provided insight on their dark matter halos using samples selected from wide-area spectroscopic surveys such as 2dF (S. Folkes et al. 1999), the Sloan Digital Sky Survey (SDSS; D. G. York et al. 2000), BOSS (I. Pâris et al. 2012), and eBOSS (A. D. Myers et al. 2015). Optically selected quasars samples have a preference for halo masses of a few times $10^{12}\ M_\odot$, and an evolving bias consistent with this halo mass scale at most redshifts (Y. Shen et al. 2007, 2009; N. P. Ross et al. 2009; S. Eftekharzadeh et al. 2015; S. A. Rodríguez-Torres et al. 2017). Remarkably, this halo mass scale coincides with the peak in the stellar-to-halo mass relation (SHMR; P. S. Behroozi et al. 2012; P. Behroozi et al. 2019), indicating a mass scale where halos are most efficient at converting gas into stars.

It is also necessary to understand the connection between AGNs and dark matter halos with consideration of the stellar masses of the AGN host galaxies (e.g., V. Allevato et al. 2019; J. Aird & A. L. Coil 2021), which can be obtained through spectral energy distribution (SED) model fitting (A. D. Goulding et al. 2018) or 2D image decomposition (e.g., J. Li et al. 2021). Recently, C. Krishnan et al. (2020) show that the host galaxies of quasars dictate the clustering of X-ray AGNs, and hence their dark matter property. They study the inferred bias for AGNs and claim that AGNs lie between matched star-forming and passive galaxy populations since their hosts are simply a mixture of the two populations.

As well, there has been much interest to probe the environmental differences between various AGN populations (e.g., obscured and unobscured). For example, the clustering of measurements of X-ray AGNs out to $z \sim 2$ shows that obscured AGNs have similar clustering properties to their unobscured counterparts (R. C. Hickox et al. 2011; L. Koutoulidis et al. 2018). These claims appear to be robust given further effort by M. A. DiPompeo et al. (2017) using a half million Wide-field Infrared Survey Explorer (WISE)-selected AGNs centered at $z = 1$. A. J. Mendez et al. (2016) expanded such studies to X-ray-, IR-, and radio-selected AGNs and also found no significant difference between obscured and unobscured AGNs while matching on stellar mass and star formation rate (SFR). Alternatively, V. Allevato et al. (2019) find a negative dependence of the large-scale bias on SFR using obscured AGNs, which may be expected since gas levels in galaxies will be suppressed in denser environments. Interestingly, N. Jiang et al. (2016) detect stronger clustering for obscured AGNs at scales smaller than 100 kpc $h^{-1}$ and M. Krumpe et al. (2018) detect stronger clustering for high X-ray luminosity AGNs also at smaller scales. Furthermore, M. C. Powell et al. (2018) claim a difference in clustering between the obscured and unobscured AGNs at the same luminosity, redshift, stellar mass, and Eddington ratio. They also study the relation between the black hole mass and galaxy properties and conclude that massive black holes tend to reside in central galaxies. In addition, Y. Shirasaki et al. (2018) find no obvious dependence of clustering on black hole mass.

To extract further information from clustering measurements, proper modeling (i.e., halo occupation distributions (HODs); Z. Zheng et al. 2007; Y. Zu & R. Mandelbaum 2015; M. A. DiPompeo et al. 2017; S. Alam et al. 2021) allows one to understand not only the halo mass, but the contribution of the one-halo term (within the virial radius), two-halo term (scale beyond the virial radius), and the satellite fraction. In recent HOD modeling efforts (e.g., Y. Zu & R. Mandelbaum 2015; S. Alam et al. 2021), various galaxy tracers have been used from the eBOSS survey (R. Ahumada et al. 2020). They find that quasars have larger satellite fraction than the other two tracers (luminous red galaxies (LRGs) and emission line galaxies (ELGs)) and no environmental preference, whereas LRGs prefer denser environments and ELGs prefer less dense environments.

Complementary to clustering measurements, galaxy–galaxy lensing is a powerful tool for the direct detection of the underlying dark components of galaxies. The SDSS (D. G. York et al. 2000) pioneered these studies (E. S. Sheldon et al. 2004; R. Mandelbaum et al. 2008; W. Luo et al. 2017, 2018) with the help of its wide area covering up to 7500 deg$^2$. R. Mandelbaum et al. (2009) measured the halo mass of optically and radio-selected AGNs with the latter based on NVSS (J. J. Condon et al. 1998) and FIRST (R. H. Becker et al. 1995). They found that the halo mass of radio-loud AGNs is more than 1 magnitude larger than their optical counterparts and twice larger than a control sample with an identical stellar mass distribution. Since the SDSS imaging is too shallow to measure the galaxy–galaxy lensing signal for high-redshift objects, deep fields such as COSMOS (N. Scoville et al. 2007) provide imaging sufficient for weak lensing studies such as those of distant AGNs. For example, A. Leauthaud et al. (2015) measured the galaxy–galaxy lensing signal around moderate-luminosity X-ray-selected AGNs with $z < 1.0$ from XMM-COSMOS (N. Cappelluti et al. 2009) and C-COSMOS (M. Elvis et al. 2009). A halo mass of $10^{12.5}\ h^{-1}\ M_{\odot}$ for X-ray AGNs is estimated by fitting a global SHMR (A. Leauthaud et al. 2012). However, the statistical significance is limited since the COSMOS field is only 2 deg$^2$ thus the AGN sample size is relatively small (385) compared to the SDSS studies.

To overcome issues with sample size, the Hyper Suprime-Cam Subaru Strategic Program (HSC-SSP; H. Aihara et al. 2018) is designed particularly for weak lensing studies with a unique combination of wide area coverage, deep depths, and superb seeing conditions. The $i$-band median seeing is $0\overset{''}{.}58$ and $5\sigma$ detections are made down to an $i$-band magnitude of 26, with a weighted source number density of 21.8 galaxies arcmin$^2$. The full-color full-depth area of the first public shape catalog S16A (R. Mandelbaum et al. 2018b) covers 136.9 deg$^2$ split in six separate regions overlapping with other surveys, i.e., XMM-Newton (G. Hasinger et al. 2001), HECTOMAP (J. Sohn et al. 2018a, 2018b), GAMA09H and GAMA15H (A. Robotham et al. 2010), WIDE12H, and VVDS (O. Le Fèvre et al. 2005), which was deliberately designed to maximize the weak lensing studies with ancillary information.

Here, we measure the galaxy–galaxy lensing signal of 48,907 AGNs ranging in redshift from 0.2 to 1.2 as identified from HSC and WISE photometry using the S16A SSP data. We take advantage of the wide and deep imaging to study the lensing signal around AGNs further split into bins of stellar mass to investigate the stellar mass to halo mass relation of luminous AGNs. Equally important, we construct a control sample of 6539,241 galaxies in HSC-SSP S16A fields by matching the 2D distribution in stellar mass and redshift. We then model the galaxy–galaxy lensing signals to extract information on the halo mass.

The structure of this paper is organized as follows. In Section 2, we present the data including the shape catalog from HSC-SSP S16A and the AGN hosts as foreground lenses. We also describe several basic properties of the foreground samples for the AGNs and control galaxies. We specify the galaxy–galaxy lensing estimator in Section 3 and our forward modeling method in Section 4. The major results are presented in Section 5. We discuss these results in Section 6 and summarize the work in Section 7. Unless stated elsewhere, we adhere to using Planck 2018 (Planck Collaboration et al. 2020) cosmology with $\Omega_m = 0.315$, $h = 0.674$, and $\sigma_8 = 0.810$.

## 2. Data

### 2.1. SUBARU HSC-SSP

The HSC-SSP (H. Aihara et al. 2018, 2019) is an optical imaging survey of over 1000 deg$^2$ in five optical bands (*grizy*) with the Subaru Telescope. The survey has three layers (Wide, Deep, and UltraDeep) with depths reaching an AB magnitude of $i \sim 26.4$, $\sim$26.5, and $\sim$27.0 mag (5$\sigma$ for point source), respectively. The Wide survey regions target areas overlapping with other surveys, i.e., VVDS, XMM, GAMA09H, GAMA15H, HECTOMAP, and WIDE12H.

The imaging data used in this paper are processed using the HSC pipeline (hscPipe 4.0.2; J. Bosch et al. 2018), which is based on the software built for LSST (M. Juric & T. Tyson 2015). The software removes instrumental effects such as flat-fielding, bias subtraction, the nonuniformity of the plate scale, cosmic rays, and bad pixels. The major systematic in weak lensing measurement is inaccuracies in the determination of the point-spread function (PSF). The modeling of the PSF at the positions of source galaxies is reconstructed using the PSFEx package developed by E. Bertin (2011). 20% of the stars are not used in the reconstruction but reserved to test the performance of the PSF reconstruction. The reconstruction of the PSF is first performed on each exposure and then coadded in a consistent manner for the coadded images.

After the PSF is reconstructed, the source galaxy shapes are then measured using the re-Gaussianization technique (R. Mandelbaum et al. 2018b) in order to generate a shape catalog of galaxies for weak lensing science. The multiplicative bias ($m$) and the additive bias ($c$) of the shape measurement $\gamma^{\rm obs} = (1+m)\gamma^{\rm true} + c$ are calibrated via a set of image simulations matching the properties of the HSC survey that are generated (R. Mandelbaum et al. 2018a) by the GalSim software (B. T. P. Rowe et al. 2015).

The final galaxy shape catalog has more than 12 million galaxies after several quality control cuts and covers a total area of 136.9 deg$^2$. The posterior distribution for the photo-$z$ for each galaxy is estimated using six different methods as described in M. Tanaka et al. (2018), i.e., the Direct Empirical Photometric code (B. C. Hsieh & H. K. C. Yee 2014), Extended Photometric Redshift,[13] Flexible Regression over Associated Neighbors with Kernel dEnsity estimatioN for Redshifts,[14] machine learning for photo-$z$ based on a self-organizing map (SOM; M. Carrasco Kind & R. J. Brunner 2014), nearest neighbors $P(z)$ (C. E. Cunha et al. 2009), and Mizuki (M. Tanaka 2015). In this paper, we use the Mizuki photometric redshifts, which also provide a stellar mass estimate for each galaxy.

### 2.2. Active Galactic Nucleus Sample

While most AGNs are intrinsically luminous in any given wave band, the homogeneous selection of an unbiased population of AGN from survey data is not straightforward (e.g., R. C. Hickox et al. 2009; A. D. Goulding et al. 2012). Intervening gas and dust, as well as dilution of the AGN signatures by host galaxy light are the most common sources of observation selection bias. Indeed, many studies have now shown that no one wave band can identify all AGNs (see P. Padovani et al. 2017 for an extensive review). Moreover, those systems undergoing mergers and are likely heavily dust obscured, further hampering AGN detections (R. C. Hickox et al. 2014; X. Chen et al. 2015; A. D. Goulding et al. 2018). However, even in the presence of significant dust and gas attenuation, relatively unbiased detections of luminous AGNs may be made when considering information in the mid-infrared regime. For example, AGN emission produced directly from the optical/UV luminous accretion disk or from the X-ray emitting corona may be absorbed and reprocessed by dust which surrounds the central black hole. This dust-rich torus isotropically reemits at mid-infrared wavelengths, which is relatively insensitive to further absorption at larger radial distances from the AGN. These obscured AGN signatures are imprinted in the galaxy-wide photometry.

When the AGN outshines the host galaxy, rigid boundaries in color–color diagrams are often used to distinguish AGNs from nonactive systems (e.g., D. Stern et al. 2005; J. L. Donley et al. 2007; R. J. Assef et al. 2013). However, these regions of single color–color space that are occupied by AGN are still contaminated by stars and/or the far more numerous nonactive systems due to variation in galaxy properties or photometric uncertainties. Substantial progress has been made by simultaneously considering the dust-independent information from WISE in the mid-infrared along with highly precise shorter wavelength information in the optical under the framework of unsupervised machine learning.

Here we utilize the quasar catalog from A. D. Goulding et al. (2024; submitted; hereafter, G24), which uses a sophisticated unsupervised dimensionality reduction technique to capture all of the multidimensional color, magnitude, and size space, and distill this down into a single interpretable 2D manifold. Using the combination of WISE W1–3 mid-infrared and Hyper Suprime-Cam *grizy* imaging along with the Uniform Manifold Approximation and Projection (UMAP) algorithm, G24 robustly identify a sample of $\sim$290,000 quasars from the $\sim$628 deg$^2$ (unmasked) region covered by HSC-SSP Wide and WISE. Briefly, UMAP incorporates the input multidimensional photometric information to construct a simplicial complex topology that groups objects with similar properties, allowing them to cluster, while simultaneously forcing objects with dissimilar properties to diverge. The result is a simple 2D space allowing AGN selection. Inputting a test sample with no prior knowledge of the intrinsic source properties, G24 show that UMAP correctly segregates known stars from galaxies, and known AGN/quasars from galaxies. G24 further provide a demarcation between obscured and unobscured AGN. In addition, G24 use this quasar catalog and the associated photometry to train an augmented random forest to compute robust photometric redshifts, which we make use of here. These machine learning–based photo-$z$s are shown to perform equally well on Type 1 and Type 2 AGNs alike to $z \sim 3$ to an average precision of $\delta z/(1+z) \sim 0.02$ and 0.03, respectively.

Here, we additionally perform SED model fits including both stellar and AGN components to compute stellar masses and SFRs for the AGN sample. To perform the SED modeling we use `X-CIGALE` (G. Yang et al. 2020), an extension to the `CIGALE` SED-fitting package (L. Ciesla et al. 2015). The main advantages to the use of `X-CIGALE` over the standard `CIGALE` SED package is the inclusion of the new `SKIRTOR` accretion models that invoke a clumpy torus structure as well as a polar dust emission component to characterize the AGN emission. For quasars with previously available spectroscopy, we use these spectral redshifts as inputs to `X-CIGALE`, while for all remaining quasars in our redshift range, we use the photo-$z$s provided in G24.

[13] https://hsc-release.mtk.nao.ac.jp/doc/index.php/photometric-redshifts/
[14] https://github.com/joshspeagle/frankenz

**Table 1**
Basic Properties of the Active Galactic Nucleus Samples and Control Galaxy Samples

| Sample | Number | $\langle z \rangle$ | $\langle \log(M_*/h^{-2}M_\odot) \rangle$ |
|---|---|---|---|
| AGN: all | 48,907 | 0.77 | 10.64 |
| AGN: Type I | 11,331 | 0.78 | 10.75 |
| AGN: Type II | 37,576 | 0.80 | 10.56 |
| AGN: $0.2 \geqslant z < 0.65$ | 11,647 | 0.54 | 10.38 |
| AGN: $0.65 \geqslant z < 0.80$ | 10,652 | 0.74 | 10.59 |
| AGN: $0.80 \geqslant z < 1.20$ | 21,475 | 0.95 | 10.86 |
| AGN: $L_{\rm AGN} \geqslant 44.7$ | 26,226 | 0.97 | 10.72 |
| AGN: $L_{\rm AGN} < 44.7$ | 22,681 | 0.57 | 10.55 |
| AGN: $M_* \leqslant 10.4$ | 12,107 | 0.70 | 10.04 |
| AGN: $10.4 < \log(M_*) \leqslant 10.8$ | 12,875 | 0.76 | 10.61 |
| AGN: $10.8 < \log(M_*) \leqslant 11.1$ | 10,646 | 0.86 | 10.95 |
| AGN: $11.1 < \log(M_*)$ | 8145 | 0.90 | 11.31 |
| Galaxy: $\log(M_*) \leqslant 10.4$ | 1,392,891 | 0.67 | 10.04 |
| Galaxy: $10.4 < \log(M_*) \leqslant 10.8$ | 278,588 | 0.74 | 10.61 |
| Galaxy: $10.8 < \log(M_*) \leqslant 11.1$ | 115,205 | 0.82 | 10.95 |
| Galaxy: $11.1 < \log(M_*)$ | 62,937 | 0.86 | 11.31 |

The number of AGNs falling in the HSC-SSP S16A region is 48,907, which are further divided into Type I and Type II, three redshift bins, as well as four stellar mass bins. Table 1 lists the mean value of the redshift and stellar mass of each subsample. The mean redshift of all AGNs increases from 0.70 to 0.90 as the stellar mass becomes larger. The AGN sample as a whole has a mean redshift of 0.77 and stellar mass of $\log(M_*/h^{-2}M_\odot) = 10.64$. Type II AGNs outnumber the Type I AGNs by a factor of 3 and have higher stellar masses as well.

### 2.3. Control Sample of Galaxies

The control sample of galaxies contains all objects from HSC-SSP Data Release 1 (DR1) with redshifts ranging from 0.2 to 1.2, similar to the AGN sample. The photometric redshift estimation of HSC-SSP DR1 (M. Tanaka et al. 2018) is based on Mizuki (M. Tanaka 2015), an SED-fitting code which uses a set of templates generated using stellar population synthesis models (G. Bruzual & S. Charlot 2003) under the assumption of a Chabrier initial mass function (G. Chabrier 2003) and the Calzetti dust attenuation (D. Calzetti et al. 2000) curve. Emission lines are added to the templates assuming solar metallicity (A. K. Inoue 2011). In addition to galaxy templates, they also include AGN templates generated by combining the Type I AGN spectrum from M. Polletta et al. (2007) and young galaxy templates from G. Bruzual & S. Charlot (2003).

We calculate the weak lensing signal using these photometric redshifts for the lens galaxies. Given that the control sample of galaxies does not have the same distributions in stellar mass and redshift (see the left and middle panels of Figure 1), we reweight the control sample of galaxies in order to match the 2D ($21 \times 21$ bins) distributions of $M_{\rm stellar}$ ($8.0 < \log M_{\rm star} < 12.5$) and redshift ($0.2 < z < 1.2$) of the AGN sample. The weight is the ratio between the number of AGNs and galaxies in each bin with a smallest value of 0.003 and a maximum weight value of 2.28. This can be easily calculated such that

$$w_l = \frac{N_{\rm qso}(z,\, M_{\rm stellar})}{N_{\rm gal}(z,\, M_{\rm stellar})}. \tag{1}$$

In total, there are 1,849,621 galaxies in this sample, so the signal-to-noise ratio (SNR) for the control sample is significantly higher than that of the AGN sample. We follow the same mass binning of the AGN samples. We then reweight the subsamples of galaxies to match the $M_*{-}z$ space of the four AGN stellar mass bins with the intention of comparing the halo masses between the galaxies and AGNs fairly. The last four rows of Table 1 list the number of galaxies in each bin as well as the mean redshift and stellar mass after reweighting. The stellar masses are identical to those of the AGN samples while the redshifts are in general lower than for the AGN samples. That is mainly due to the fact that there are much fewer high-redshift galaxies that have stellar mass measurements.

## 3. Excess Surface Density Estimator and Background Source Selection

We describe the procedure to measure the galaxy–galaxy lensing signal using faint galaxies imaged by HSC including corrections for known systematics.

### 3.1. Background Sources

The shape catalog (R. Mandelbaum et al. 2018b) is based on $i$-band coadded images with the re-Gaussianization method (C. Hirata & U. Seljak 2003). The requirement of full-depth and full-color imaging results in an area coverage of 136.9 deg$^2$, which ensures uniformity of the galaxy number density. As in R. Mandelbaum et al. (2018b), we limit the cmodel magnitude (see the definition in J. Bosch et al. 2018) to $i_{\rm cmodel} < 24.5$.

The measurement biases for each galaxy are calibrated based on the simulations generated by the software GalSim (B. T. P. Rowe et al. 2015). These biases include multiplicative bias $m$, additive bias ($c_1$, $c_2$), shape measurement error $\sigma_e$ and shape noise $e_{\rm rms}$ as a function of SNR and resolution. In the simulations, realistic images from the Hubble Space Telescope Advanced Camera for Surveys F814W images of the COSMOS field (A. Leauthaud et al. 2007) are used to generate galaxies with various morphologies. The overall systematic uncertainty is around 1% (R. Mandelbaum et al. 2018a). As mentioned in Section 2.1, we choose the Mizuki photometric redshift catalog, which is the only one that contains stellar mass information among the six photometric catalogs.

### 3.2. Estimator

The galaxy–galaxy lensing signal measures the differential profile of the projected mass density (i.e., the excess surface density (ESD))

$$\Delta\Sigma(R) = \bar{\Sigma}(\leqslant R) - \Sigma(R), \tag{2}$$

where $\bar{\Sigma}(\leqslant R)$ is the average surface density averaged inside radius $R$ and $\Sigma(R)$ is the surface density at radius $R$. The signal around each galaxy sample can be measured by stacking the background shapes of galaxies (H. Miyatake et al. 2019)

$$\Delta\Sigma(R) = \frac{1}{2\mathcal{R}(R)} \frac{\sum_l^{N_l} w_l \sum_s^{N_s} w_{ls} e_{t,ls} [\langle \Sigma_{cr}^{-1} \rangle_{ls}]^{-1}}{[1 + K(R)] \sum_l^{N_l} w_l \sum_s^{N_s} w_{ls}}, \tag{3}$$

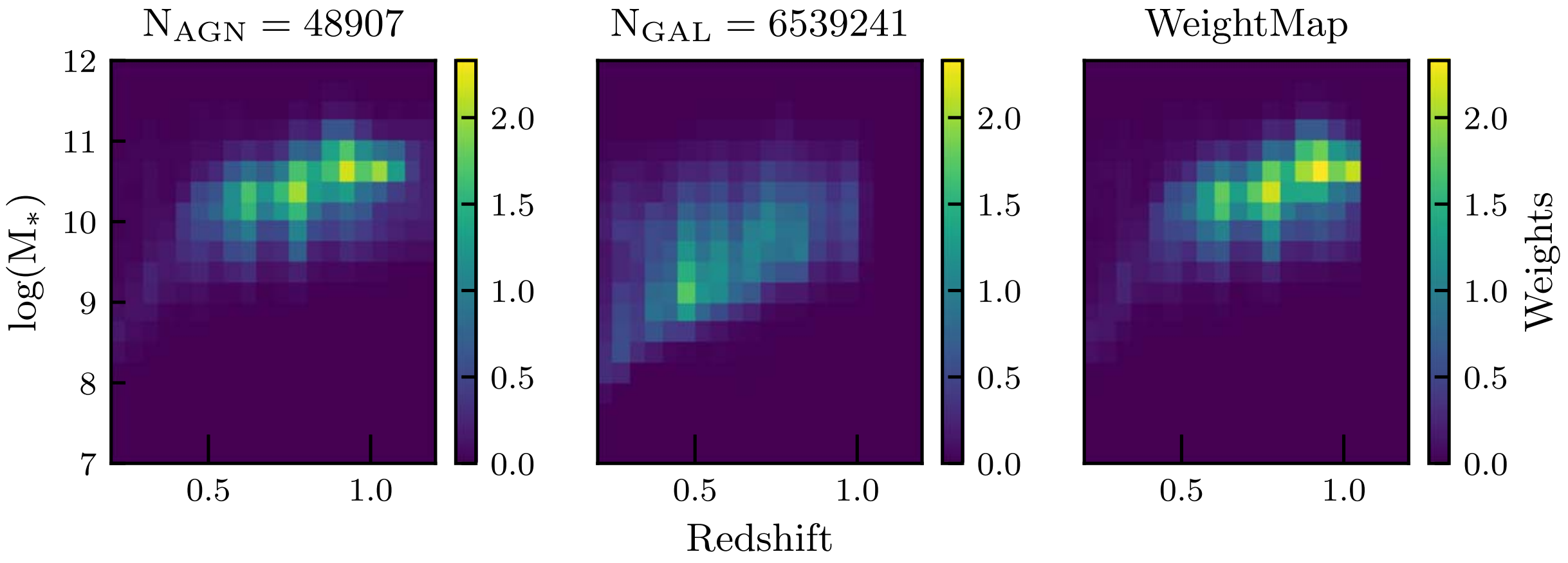

**Figure 1.** 2D density plot of stellar mass and redshift with the corresponding weight map (Left: AGNs. Middle: control galaxy sample. Right: weight map). Each panel has 21 × 21 bins with a minimum weight value of 0.003 and a maximum value of 2.28.

where

1. $e_{t,ls}$ is the tangential component of a source galaxy that is within projected distance $R$ away from a lens galaxy.
2. $\mathcal{R}(R) = 1 - \frac{\sum_l^{N_l} w_l \sum_s^{N_s} e_{rms}^2 w_{ls}}{\sum_l^{N_l} w_l \sum_s^{N_s} w_{ls}}$ is the responsivity of shape estimator, which is ≈0.84.
3. $w_{ls}$ is the weight for each source galaxy, $w_{ls}$= $(\langle \Sigma_{cr}^{-1} \rangle_{ls})^2 \frac{1}{\sigma_e^2 + e_{\rm rms}^2}$, as a function of $\sigma_e$, $e_{\rm rms}$, and a geometric factor $(\langle \Sigma_{cr}^{-1} \rangle_{ls})^2$. ($\sigma_e$ and $e_{\rm rms}$ are the shape noise and shape measurement error, respectively, and $N_l$ and $N_s$ are the stacked number of lens galaxies and source galaxies, respectively.)
4. $w_l$ is a weight for the lens sample. For quasars we chose a weight equal to unity, while for the control sample of galaxies, the weight is given by Equation (1).
5. $\langle \Sigma_{cr}^{-1} \rangle_{ls}$ is calculated for each lens–source pair by applying the $P(z)$ of the source to marginalize over the photometric redshift errors

$$\langle \Sigma_{cr}^{-1} \rangle_{ls} = \frac{\int_{z_l}^{\infty} \Sigma_{cr}^{-1}(z_l, z) P(z) dz}{\int_0^{\infty} P(z) dz}. \tag{4}$$

6. The factor $K(R)$ accounts for the multiplicative bias calibrated from simulations

$$K(R) = \frac{\sum_l^{N_l} w_l \sum_s^{N_s} m_s w_{ls}}{\sum_l^{N_l} w_l \sum_s w_{ls}}. \tag{5}$$

Finally, we only select sources behind each lens that satisfy the requirement in E. Medezinski et al. (2018) where the accumulated probability of the photometric redshift $P(z \geqslant z_l + 0.2)$ of each source is larger than 0.98

$$P(z \geqslant z_l + 0.2) = \int_{z_l+0.2}^{\infty} p(z) dz \geqslant 0.98. \tag{6}$$

When measuring the lensing signals, we divide the projected radius into 10 equal logarithmic bins from 0.02 to 2 Mpc $hr^{-1}$.

#### 3.2.1. Covariance Matrix

We compute the covariance matrix from a bootstrap sample by dividing the sky using HEALPix[15] (K. M. Górski et al. 2005) into 43 subregions, which balances the number of subsamples and the size that our measurements can reach (2 Mpc $h^{-1}$) and then carry out a bootstrap routine (2000 times) to determine the covariance matrix. At small scales within the virial radius of the halos, the dominant error is the shape noise; at larger scales it has correlations caused by large-scale structure. The bootstrap sampling can capture all those features in a covariance matrix. However, due to the smaller sample size, the covariance matrix is pretty noisy. We follow the treatment in A. Leauthaud et al. (2017) that truncates the covariance matrix with 0.2 as a threshold. Figure 2 elucidates the covariance matrix.

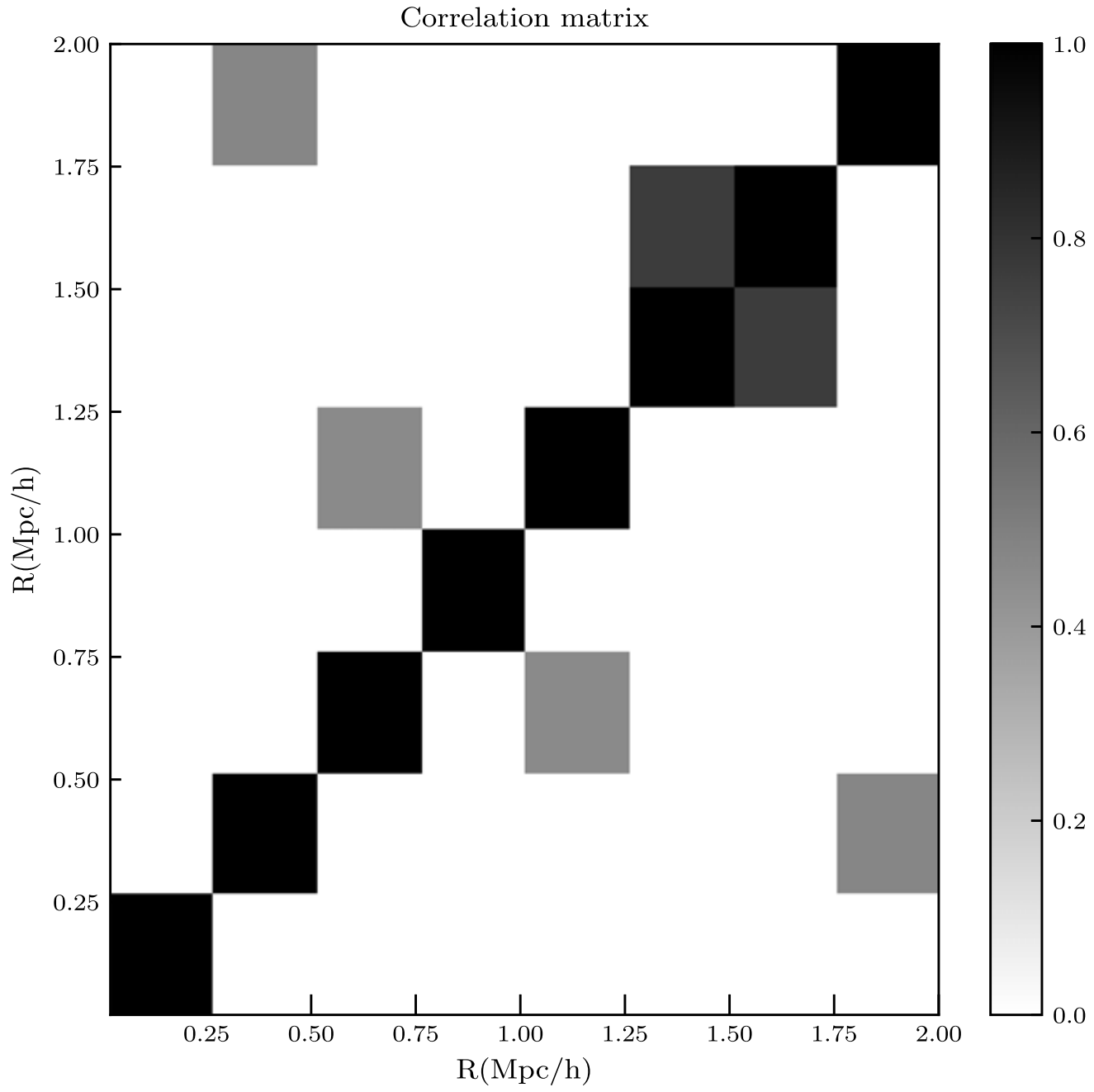

**Figure 2.** The covariance matrix calculated using 2000 bootstrap of 43 regions from S16A area divided by HEALPix.

#### 3.2.2. Photometric Redshift Bias Correction

We use H. Miyatake et al. (2019)'s method to estimate the bias from photometric redshift errors, which is redshift

15 https://healpix.sourceforge.io/

dependent

$$1 + b(z_l) = \frac{\Delta\Sigma}{\Delta\Sigma^{\rm true}}(z_l) = \frac{\sum_{ls} w_{ls}{}^{*} \langle \Sigma_{cr,ls}^{-1} \rangle^{-1} [\Sigma_{cr,zl}^{\rm true}]^{-1}}{\sum_{ls} w_{ls}{}^{*}}. \quad (7)$$

The superscript true represents the "true" value of the geometric factor using the COSMOS region overlapping with HSC-SSP S16A, and the weight $w_{ls}{}^{*}$ is provided in the PDR2 release based on the SOM technique. We follow H. Miyatake et al. (2019) to use $w_{\rm SOM}^{*} w_{ls}$ to calculate the bias in Equation (7), such that

$$w_{ls}{}^{*} = D_l^{-2}(1 + z_l)^{-2} \sum_s w_{\rm som} w_{ls}, \quad (8)$$

where the $w_{\rm SOM}$ is the weight generated based SOM (S. More et al. 2024, in preparation) and $w_{ls}$ is the weight before multiplying by $w_{\rm SOM}$. The final photometric redshift bias for each sample is then averaged over the redshift distribution as in Equation (22) of R. Nakajima et al. (2012), where

$$\langle b_{{\rm photo}-z} \rangle = \frac{\int dz p(z_l) w_{ls}{}^{*} b(z_l)}{\int dz p(z_l) w_{ls}{}^{*}}. \quad (9)$$

Even though our sample is at relatively high redshift the photo-$z$ bias is low roughly about 2%, which is consistent with H. Miyatake et al. (2019) and much smaller than the statistical error; a correction of $(1+b_{\text{photo-}z})$ is applied to the ESDs to correct for this bias.

## 4. Models

In this section, we present the modeling scheme of the galaxy–galaxy lensing signals. In general we adopt the model of R. Mandelbaum et al. (2009) to extract the information of host halo mass, halo concentration, and subhalo fraction. We fix the amplitude of the satellite HOD function because only the shape of the function matters after normalization in Equation (17). We decompose the ESD into four components, the contribution of stellar mass at small scales, the subhalo and the host halo contributions, and the two-halo term

$$\Delta\Sigma(R) = \Delta\Sigma_{\rm stellar}(R) + (1 - f_{\rm sat})\Delta\Sigma_{\rm cen}(R) + f_{\rm sat}\,\Delta\Sigma_{\rm sat}(R) + \Delta\Sigma_{\rm 2halo}(R). \quad (10)$$

We take the mean stellar mass of the lens samples as a point mass contributing to the ESD (W. Luo et al. 2018), due to the fact that the size of the galaxy is smaller than the scales relevant to our measurements, such that

$$\Delta\Sigma_{\rm stellar}(R) = \frac{\langle M_{\rm stellar} \rangle}{\pi R^2}. \quad (11)$$

Here, $\Delta\Sigma_{\rm cen}$ is the contribution of the halo given that the AGN or galaxy is located at the center of the halo. We use X. Yang et al. (2006)'s formulation to model the ESD, which assumes a Navarro–Frenk–White (NFW) density profile

$$\rho(r) = \frac{\rho_0}{(r/r_s)(1 + r/r_s)^2}, \quad (12)$$

with $\rho_0 = \frac{\bar{\rho}\Delta_{\rm vir}}{3I}$, where $\Delta_{\rm vir} = 200$ and $I = \frac{1}{c^3}\int_0^c \frac{xdx}{(1+x)^2}$. Here $c$ is the concentration parameter defined as the ratio between the virial radius of a halo and its characteristic scale radius $r_s$. The projected surface density then can be analytically expressed (X. Yang et al. 2006) as

$$\Delta\Sigma_{\rm cen}(R) = \frac{M_h}{2\pi r_s^2 I}[g(x) - f(x)], \quad (13)$$

$$f(x) = \begin{cases} \frac{1}{x^2-1}\left[1 - \frac{\ln\frac{1+\sqrt{1-x^2}}{x}}{\sqrt{1-x^2}}\right] & x < 1, \\ \frac{1}{3} & x = 1, \\ \frac{1}{x^2-1}\left[1 - \frac{\tan^{-1}(\sqrt{x^2-1})}{\sqrt{x^2-1}}\right] & x > 1, \end{cases} \quad (14)$$

and

$$g(x) = \begin{cases} \frac{1}{x^2}\left[\ln(x/2) + \frac{\ln\frac{1+\sqrt{1-x^2}}{x}}{\sqrt{1-x^2}}\right] & x < 1, \\ 2 + 2\ln\left(\frac{1}{2}\right) & x = 1, \\ \frac{1}{x^2}\left[ln(x/2) + \frac{\tan^{-1}(\sqrt{x^2-1})}{\sqrt{x^2-1}}\right] & x > 1, \end{cases} \quad (15)$$

where $x$ is the projected radius in units of $r_{\rm s}$.

We model the satellite contribution as an off-centered host halo, for a fraction, $f_{\rm sat}$ of our samples. This effect can be simply treated as follows. The projected surface density will change from an NFW profile $\Sigma_{\rm NFW}(R)$ to

$$\Sigma_{\rm off}(R|R_{\rm off}) = \frac{1}{2\pi}\int_0^{2\pi} \Sigma_{\rm NFW}(\sqrt{R^2 + R_{\rm off}^2 + 2R_{\rm off}R\cos\theta})\, d\theta. \quad (16)$$

Here, we assume that the satellite galaxies are distributed in the same manner as the dark matter distribution, i.e., an NFW profile (see also A. Leauthaud et al. 2015). The resulting projected density profile is then the convolution between $P(R_{\rm off}|M_h)$ and $\Sigma_{\rm host}(R|R_{\rm off})$

$$\Sigma_{\rm sat}(R) = \int_0^{\infty} n(M_h)\langle N_{\rm sat}\rangle(M_h) dM_h \times \int dR_{\rm off} P(R_{\rm off}|M_h)\Sigma_{\rm off}(R|R_{\rm off}, M_h), \quad (17)$$

where $P(R_{\rm off}|M_h)$ can analytically be expressed with the help of $f(x)$ in Equation (14). The quantity $\langle N_{\rm sat}\rangle(M_h)$ is the occupation function of satellite galaxies given a halo mass $M_h$ and $n(M_h)$ is the halo mass function based on J. L. Tinker et al. (2005). We model the satellite HOD function following Equation (12) in R. Mandelbaum et al. (2005), where

$$\langle N_{\rm sat}\rangle(M_h) = \Theta^H(M_h - M_{\rm min})\left(\frac{M_h - M_{\rm min}}{M'}\right)^{\alpha}, \quad (18)$$

and where $\Theta^H(M_h - M_{\rm min})$ is the Heaviside step function to set up the threshold with $M_{\rm min} = 3M_{\rm cen}$ ($M_{\rm cen}$ denotes the central halo mass), which we set to be 3 times the halo mass of the one for the $\Delta\Sigma(R)_{\rm cen}$ as in R. Mandelbaum et al. (2005), which is used to be the threshold of major/minor merging systems. We also fix $\alpha = 1$ following R. Mandelbaum et al. (2005). The difference between R. Mandelbaum et al. (2005) and our modeling is that we set the satellite fraction as a free parameter, while they fix it to be 0.2 as well as fixing $M'$ to be a constant. So in total, there are three parameters in our model: the central halo mass, halo concentration, and the satellite fraction. This parameterization is sufficient given the current SNR.

The mean density inside projected radius $R$ then can be obtained via

$$\Sigma(\leqslant R) = \frac{2}{R^2}\int_0^R y\, dy\, \Sigma(y|R_{\rm sig}), \tag{19}$$

and the final ESD is again

$$\Delta\Sigma(R) = \Sigma(\leqslant R) - \Sigma(R). \tag{20}$$

For the two-halo term, we first generate the power spectrum at the mean redshift of each sample using `pyCamb` (A. Lewis 2013), and then convert it to correlation function as in W. Luo et al. (2018). The bias model is taken from J. L. Tinker et al. (2010)

$$\xi_{hm} = \langle b_h\rangle \eta \xi_{mm}, \tag{21}$$

where

$$\eta(r) = \frac{(1 + 1.17\xi_{mm}(r))^{1.49}}{(1 + 0.69\xi_{mm}(r))^{2.09}}. \tag{22}$$

And the halo bias term $\langle b_h\rangle$ is the effective bias (R. Mandelbaum et al. 2005) so that

$$\langle b_h\rangle = (1.0 - f_{\rm sat})b_h(M_{\rm cen}) + f_{\rm sat}\int_0^\infty n(M_h)\langle N_{\rm sat}\rangle(M_h)b_h(M_h)dM_h. \tag{23}$$

The ESD from the two-halo term is then calculated from the matter–matter correlation given a bias term

$$\Sigma(R) = 2\bar{\rho}\int_R^\infty [1 + \langle b_h\rangle\xi_{\rm mm}(r)]\frac{r\, dr}{\sqrt{r^2 - R^2}}, \tag{24}$$

and

$$\Sigma(\leqslant R) = \frac{4\bar{\rho}}{R^2}\int_0^R y\, dy\int_y^\infty [1 + \langle b_h\rangle\xi_{\rm mm}(r)]\frac{r\, dr}{\sqrt{r^2 - y^2}}, \tag{25}$$

so the two-halo term is

$$\Delta\Sigma(R)_{\rm 2halo} = \Sigma(\leqslant R) - \Sigma(R). \tag{26}$$

Finally, we build the likelihood function and run Markov Chain Monte Carlo (MCMC) to estimate the posterior using `emcee`,[16] which implements the affine variant sampler of J. Goodman & J. Weare (2010). We assume a Gaussian likelihood with

$$\ln\mathcal{L} = -0.5[\boldsymbol{X}^T\boldsymbol{C}^{-1}\boldsymbol{X}], \tag{27}$$

where $\boldsymbol{X} = \boldsymbol{D} - \boldsymbol{Model}$ is the difference between the data vector and model, $\boldsymbol{X}^T$ denotes the transpose of $\boldsymbol{X}$, and $\boldsymbol{C}^{-1}$ is the inverse of the covariance matrix.

[16] https://emcee.readthedocs.io/en/stable/

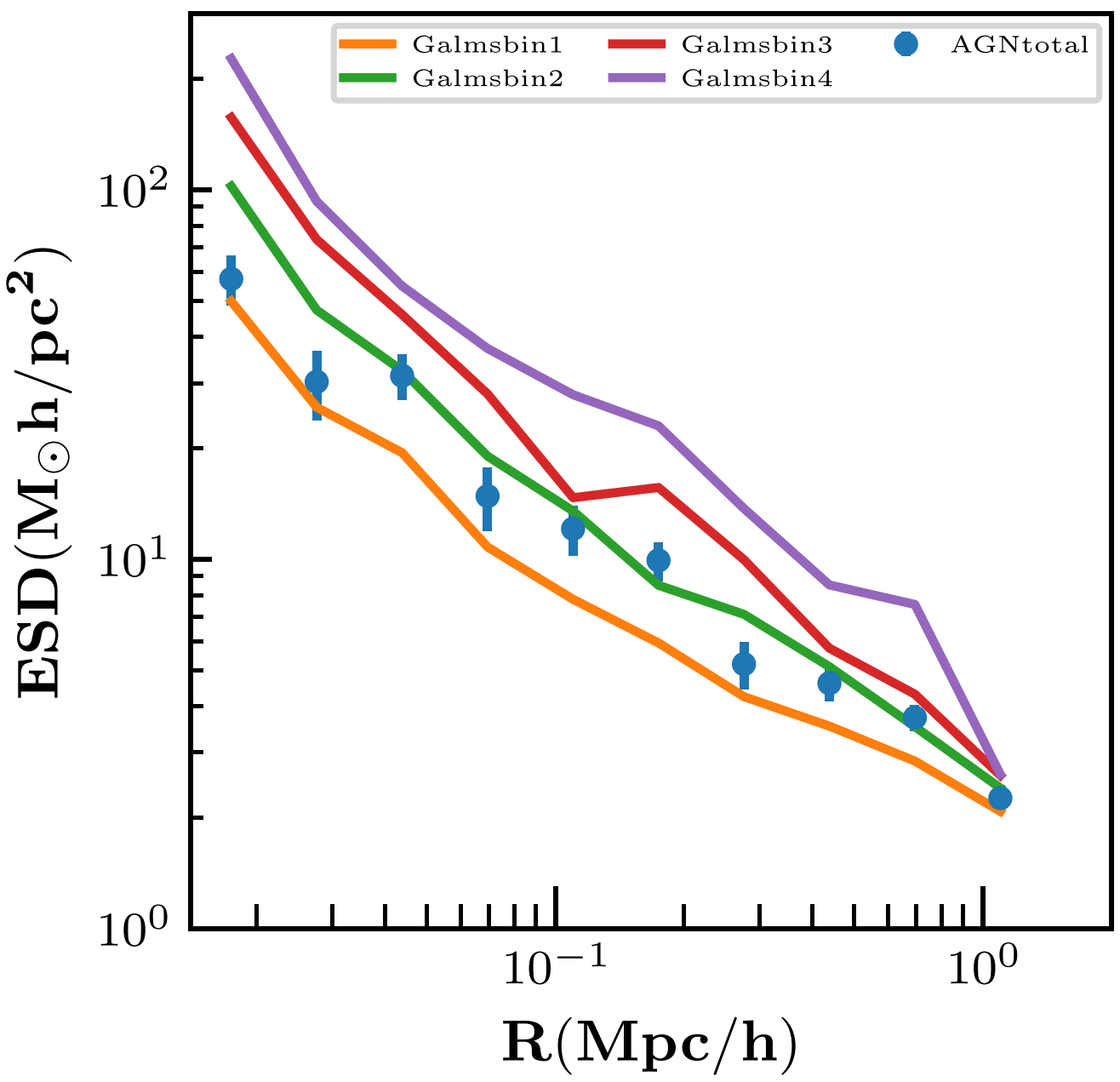

**Figure 3.** Lensing mass profile (ESD) for the full AGN sample (black circles) and the weighted control galaxy sample binned in stellar mass. The digits in the legend are the stellar mass range in $\log(M_*/h^{-2}M_\odot)$.

## 5. Results

We present the galaxy–galaxy lensing measurements of AGNs, the posterior probability distribution of the parameters in our model fit, and the SHMR of AGNs relative to the inactive galaxy population.

### 5.1. Weak Lensing Signal

In Figure 3, we show the galaxy–galaxy lensing signal measurement for the full AGN sample using the HSC-SSP S16A shape catalog. Given the AGN sample size, we are able to measure the lensing signal in 10 spatial bins. We find a highly significant lensing signal that drops off smoothly out to 2 Mpc $hr^{-1}$.

For comparison, the ESDs of a large inactive sample of galaxies are displayed in color, split into four different stellar mass bins. The ESDs are weighted according to the map of the stellar mass–redshift 2D distribution. As expected, the strength of the ESD signals for galaxies increases as a function of stellar mass. At smaller scales, the ESD of the AGN sample lies between the lensing profile of two least massive galaxy bins ($\log(M_*/M_\odot) = 10.04$–$10.61$). At larger scales (>300 kpc $hr^{-1}$) the ESD is consistent with the green curve, which belongs to the second stellar mass bin for the galaxy sample. This is mainly caused by differences in the satellite fractions. For lower stellar mass halos of the galaxy control sample, the satellite fraction increases thus causing a flattening of the ESD at larger scales. Such behavior is not evident in the AGN sample (Figure 3). These effects at these scales (∼0.3–1 Mpc $hr^{-1}$) are not likely caused by the two-halo term. We elaborate further on this in Section 5.3.

We then plot in Figure 4 the ESDs of the AGN sample split into four bins of stellar mass of their host galaxies with similar number of lenses, and hence SNR of the lensing signals. The galaxy control sample is shown for comparison using the same mass bins, which result in having the same mean redshift and stellar mass as their AGN counterparts after weighting. We find that the ESD of high mass galaxies (control sample) is significantly higher than AGNs with the same stellar mass hosts. This means that the halo masses of the control sample are much larger than their AGN counterparts of the same stellar mass. However, the AGN and control galaxies, matched at lower stellar mass, are nearly equivalent thus sharing the same environments.

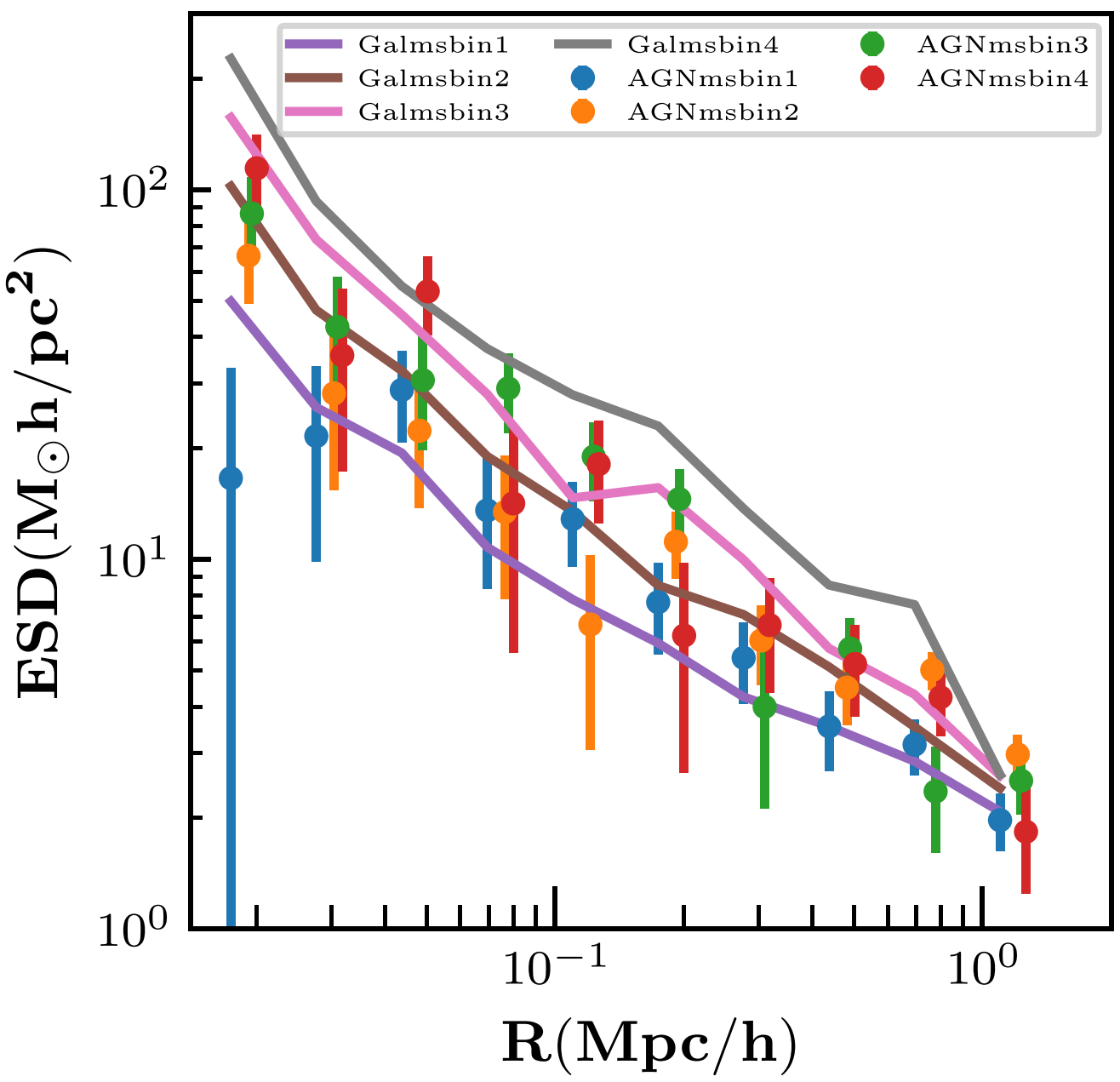

**Figure 4.** Lensing mass profile (ESD) of AGNs split by the stellar mass of their host galaxy as indicated.

### *5.2. Lensing Signal in the Active Galactic Nucleus Subsamples*

We further divide the AGN sample by their properties including AGN luminosity, type (obscured versus unobscured), and redshift. The left panel of Figure 5 compares the ESDs of AGNs of Type I (unobscured; solid dots) and Type II (obscured; empty circles). The ESDs of Type II AGNs are similar to the Type I AGNs at all scales except that the SNR of the ESD from the Type I sample is lower due to fewer lenses therein. This results in a similar halo mass estimation for Type I and II AGNs with the fitted halo mass for Type II AGNs being 0.17 dex higher than for the Type I AGNs. However, the uncertainty is larger than the difference between the two classes of AGNs given the SNR. We do not rule out the possibility that there is a real difference that needs to be further investigated. We reserve such analysis for a future study since this is not the main focus of this work.

That is also the case for the other two parameters (AGN luminosity and redshift), i.e., we find no statistical difference between the subsamples. We note that the uncertainty in the photometric redshifts may wash out any weak redshift and luminosity dependence of the mass profiles. These dependencies can be further explored with the larger survey area ($\sim 4\times$) of the HSC shape catalog in S19A (X. Li et al. 2022). Even so, we report on their difference in terms of halo mass in the Discussion.

### *5.3. Halo Mass Profile Modeling*

Using the model ESDs as described in Section 4, we decompose the contribution of the ESDs into different components: the central galaxy as an NFW profile, satellite galaxies, stellar mass of the host galaxy, and the two-halo term. We fix the stellar mass as described in Section 4, Equations (11) and (26). In Figure 6 and the upper left panel of Figure 7, we show the different contributions to the model fit for the AGNs and control sample respectively. The orange solid lines are the total signal combining all contributions. As shown, the stellar component (red solid line) only contributes at small scales (<500 kpc $hr^{-1}$). The host halo is the dominant term (solid blue curve). The satellite contribution (solid green curve) is shown at relatively large scales within the virial radius. As mentioned earlier, a larger satellite fraction leads to an upturn at larger scales and flattens the overall profile.

First, we demonstrate the accuracy of the model fits to the ESD of the control galaxy sample in Figure 7 using four bins of stellar mass. The best-fit parameters are listed in Table 2 with $1\sigma$ errors from the posterior distribution of the MCMC process.

Based on our modeling procedure (Section 4), the AGNs typically reside in halos with a characteristic mass of $\log(M_h/h^{-1}M_\odot) = 12.06$ (Table 2). The halo masses of the AGNs do not change significantly as a function of stellar mass, whereas the control samples follow the SHMR calibrated in A. Leauthaud et al. (2010). Especially in the most massive stellar mass bin, the AGN halo mass is 0.8 dex lower than the control sample.

### *5.4. Stellar-to-halo Mass Relation of Active Galactic Nuclei*

With the halo mass estimated from both the AGN and galaxy (control) samples, we measure the SHMR and compare with relations from published observations. In Figure 8, the two solid lines are from A. Leauthaud et al. (2010). The blue and orange colors represent high (0.74–1.0) and low (0.48–0.74) redshift bins. The results for the control galaxy samples are shown as the green (as six stellar mass bin) and red (as four stellar mass bin) data points. The halo mass of the total AGN sample is denoted as a red data point, and the purple dots are from the four stellar mass AGN subsamples.

Overall, the SHMR of our galaxy control sample follows the trends of both calibrated relations. However, this is not exactly the case for the halo masses of our AGN sample. The halo masses do not show as strong of a stellar mass dependence as that of the control sample. The four stellar mass bin samples roughly have similar (log) halo masses between 11.95 and 12.24. Especially the most massive stellar mass bin, its halo mass is 0.8 dex away from that of the control sample indicating that most of its gas has been converted to stars. In L. Posti et al. (2019), they found that massive spiral galaxies have $f_* = \frac{M_*}{(\Omega_b/\Omega_d)M_h}$. The halo masses of the AGNs as a whole are located above and below the knee of the SHMR, which is considered to be the most efficient "sweet spot" for star formation and gas accretion (H. Mo et al. 2010).

## 6. Discussion

The AGN–halo connection has been extensively studied in the literature. For example, C. Li et al. (2006) claim that the halos of narrow-line AGNs have a mass range from $10^{12}$ to $10^{13}\ h^{-1}\ M_\odot$, roughly spanning the peak in the SHMR, based on the projected cross correlation between AGNs and a reference galaxy sample from SDSS Data Release 4. By combining clustering and galaxy–galaxy lensing, R. Mandelbaum et al. (2009) reported that radio-loud AGNs reside in halo masses of $1.6\times10^{13}\ h^{-1}\ M_\odot$, which is twice as massive than their radio-quiet counterparts at $8 \times 10^{12}\ h^{-1}\ M_{\odot}$. Building on this, Z. Zhang et al. (2021) used the SDSS Data Release 7 shape catalog (W. Luo et al. 2017) to improve halo estimation by modeling both galaxy–galaxy lensing signals and the ratios among projected cross-correlation samples containing mostly central Type II AGNs due to their selection method; the halo mass of their full (irrespective of their radio loudness) AGN sample is $\log(M_h/h^{-1}M_{\odot}) = 11.98^{+0.06}_{-0.07}$.

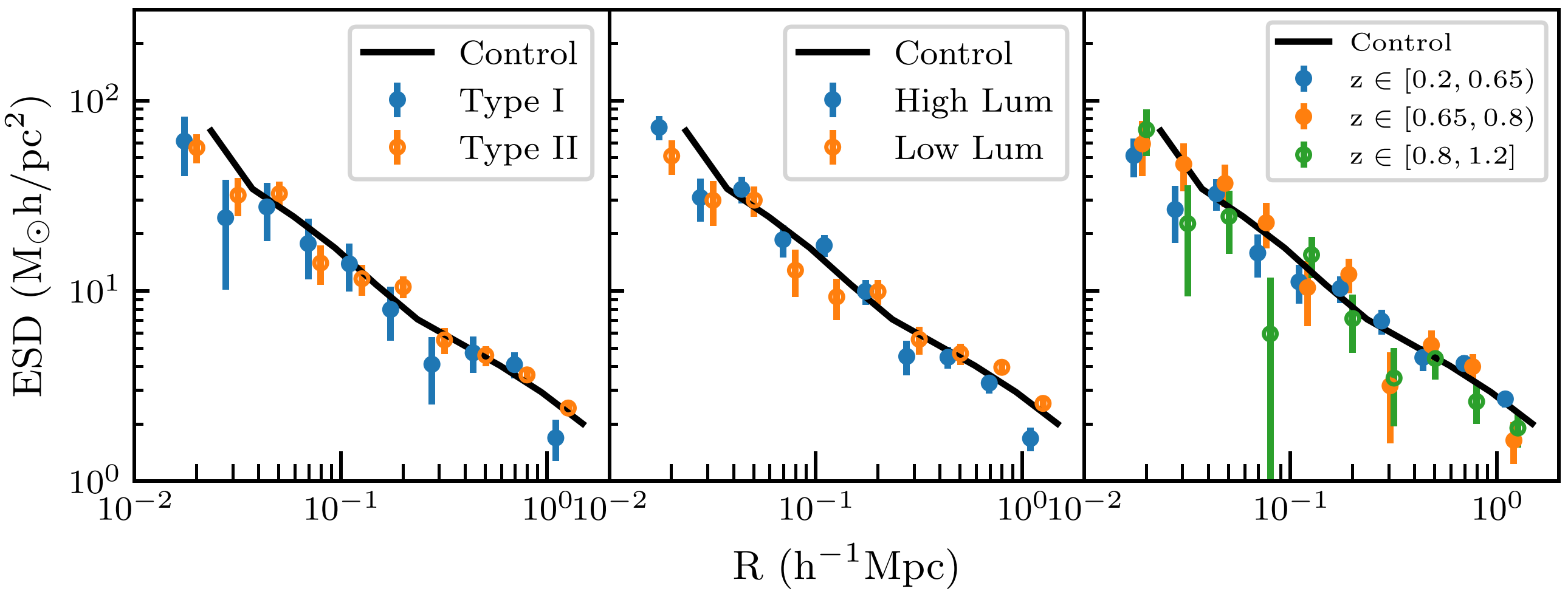

**Figure 5.** Galaxy–galaxy lensing signal (ESD) of the AGNs split by type (left), luminosity (middle), and redshift (right). The black solid lines are the results from the control sample as reference.

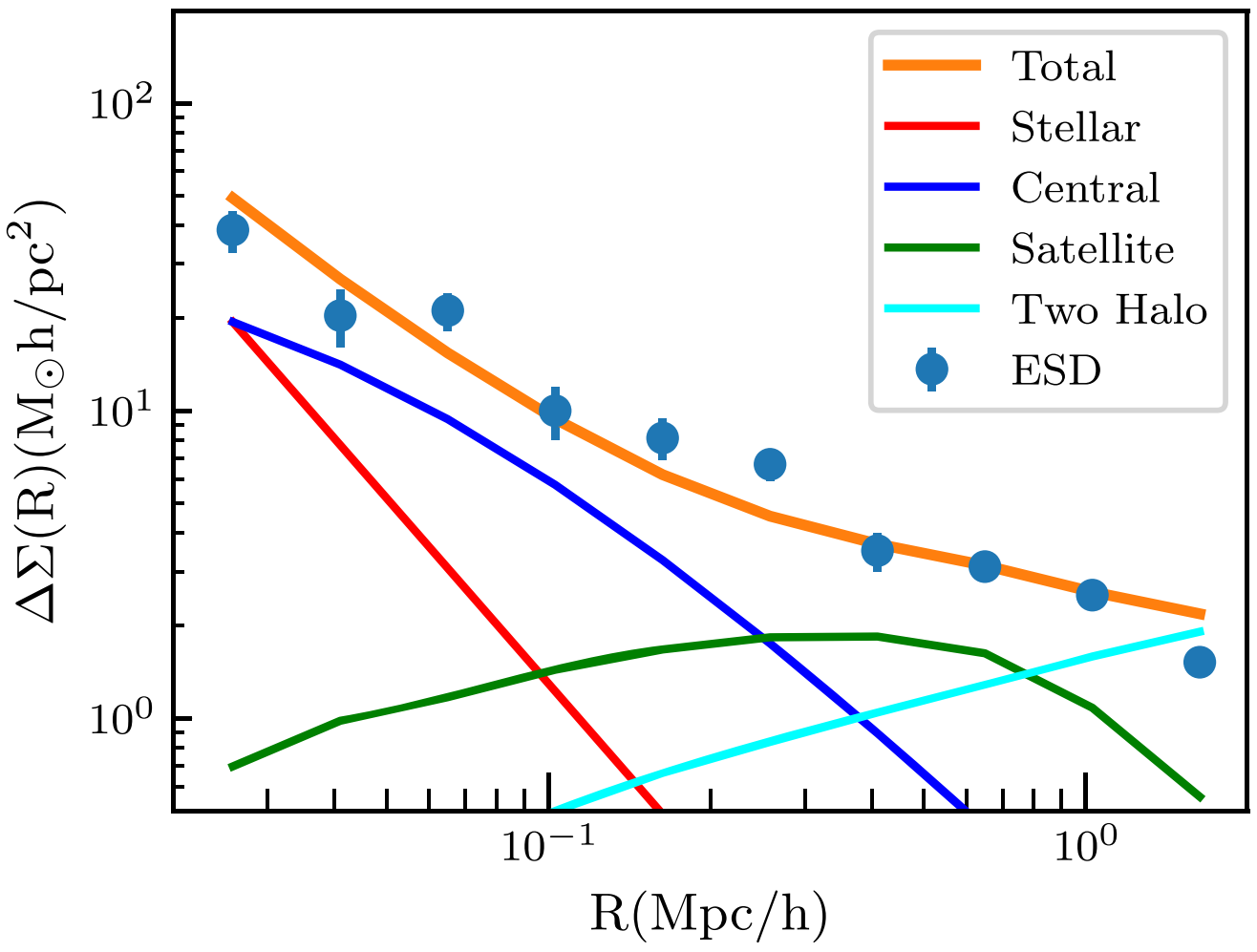

**Figure 6.** Modeling the lensing mass profile (ESD) of all AGNs as a whole. We plot the different contributions to the ESDs using different colors, i.e., stellar mass (red solid line), the central galaxy with an NFW profile (blue), satellite contribution (green), and two-halo term (cyan).

Using an X-ray-selected sample, A. Leauthaud et al. (2015) extract halo masses using the galaxy–galaxy lensing technique for 382 moderate-luminosity AGNs at $z < 1$, including both obscured and unobscured cases, from the 2 deg$^2$ COSMOS field. The halo mass is $\log(M_h/h^{-1}M_{\odot}) = 12.5$, comparable to optically selected AGN samples. Likely more relevant to our HSC study, R. C. Hickox et al. (2011) presented the clustering property of mid-infrared-selected AGNs, using a sample in the redshift range $0.7 < z < 1.8$ selected from the 9 deg$^2$ Boötes multiwavelength survey. They found that the halo masses of Type II (obscured) QSOs are larger than those of Type I (unobscured) QSOs with $\log(M_h/h^{-1}M_{\odot}) = 13.3$ and 12.7, respectively, which may pose a challenge to unified models of AGN viewing angles or represent an earlier obscured growth phase with a longer duty cycle than the unobscured state.

Our effort based on galaxy–galaxy lensing using HSC finds that the characteristic halo mass of optical + mid-infrared-selected (HSC and WISE) AGNs is $\log(M_h/h^{-1}M_{\odot}) = 12.06^{+0.11}_{-0.12}$, right at the knee of the SHMR. For a proper comparison, we reweight the control galaxy samples using the $M_*$–redshift distribution of the high- and low-luminosity AGN samples. The halo mass of the low-luminosity AGN sample is $\log(M_h/h^{-1}M_{\odot}) = 12.13^{+0.13}_{-0.15}$, which is indistinguishable from its high luminosity counterpart within the $1\sigma$ error ($\log(Mh) = 11.87^{+0.17}_{-0.18}$). This can be already expected from the ESD comparison from the middle panel of Figure 5.

The mean stellar masses of Type I AGNs are larger than those of Type II AGNs, which disagrees with the results from F. Zou et al. (2019). This may be due to the different classification methods: the latter uses the FWHM > 2000 km s$^{-1}$ of at least one emission line criterion to define Type I AGNs and the rest are Type II for X-ray-detected points sources cross identified in the Chandra Cosmos Legacy Survey. A. D. Goulding et al. (2018) use a UV/optical color cut assuming that the characteristic tail of the AGN accretion disk is absent for Type II AGNs. This selection also leads to a difference in the stellar mass distribution. In order to take selection effects into account in the galaxy–galaxy lensing measurement, we repeat the reweighting as done for the high- and low-luminosity samples. After reweighting, the typical halo mass of Type I AGNs decreases to $\log(M_h/h^{-1}M_{\odot}) = 11.85^{0.28}_{-0.30}$, which is consistent with the typical halo mass of Type II AGNs ($\log(M_h/h^{-1}M_{\odot}) = 12.02^{0.13}_{-0.13}$). The small error bar of the Type II sample is due to the larger number of Type II AGNs.

Finally, similar to the clustering results, we do not observe any redshift dependence of the halo mass of our AGN sample The lensing profiles in three different redshift bins are indistinguishable within the $1\sigma$ errors even though the highest redshift bin sample has a lower halo mass, but larger error bars due to the dearth of stacked background source galaxies. This is in agreement with J. Aird & A. L. Coil (2021), who apply a semianalytical approach to populate AGNs in the UNIVERSEMACHINE catalog (P. Behroozi et al. 2019). They found that the halo masses of AGNs have no redshift dependence for AGNs with $L_X > 10^{42}$ ergs s$^{-1}$ whereas the halo masses of the control galaxies increase by about 0.3 dex from redshifts above 2 to 0.2. They also found no $L_X$ dependence on the halo mass at redshift 0.75. The halo masses of high and low AGN luminosity samples are not distinguishable at the 1$\sigma$ level after reweighting the high luminosity sample. We will further explore such issues based on the future HSC-SSP shape catalog (X. Li et al. 2022) to pin down this difference.

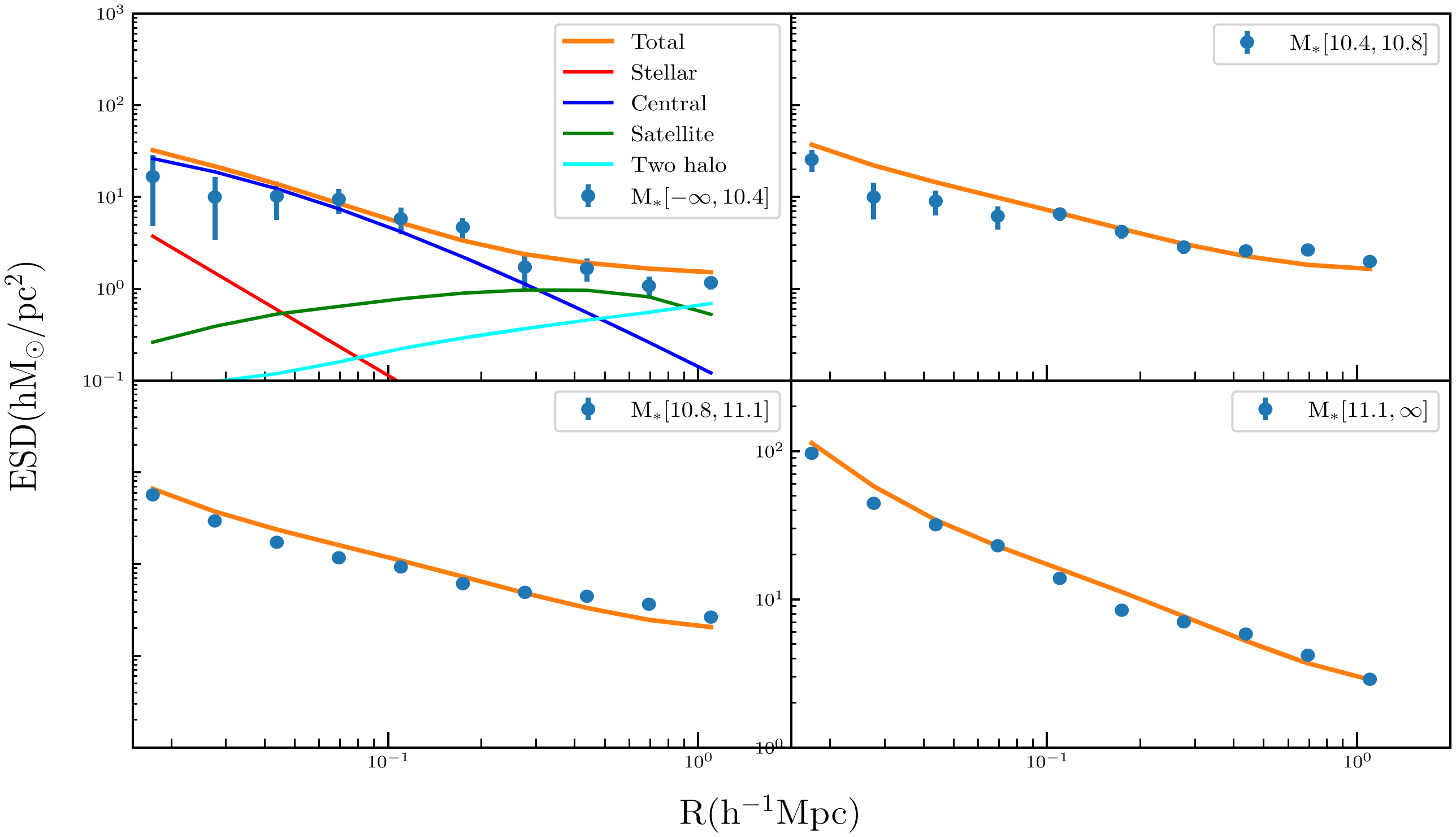

**Figure 7.** Modeling the lensing mass profile (ESD) of the control galaxies split by the stellar mass. We plot the different contributions to the ESDs in the upper left panel using different colors, i.e., stellar mass (red solid line) and the central galaxy with an NFW profile (blue). $M_*$ here denotes $\log(M_{\rm star}/h^{-2}M_\odot)$.

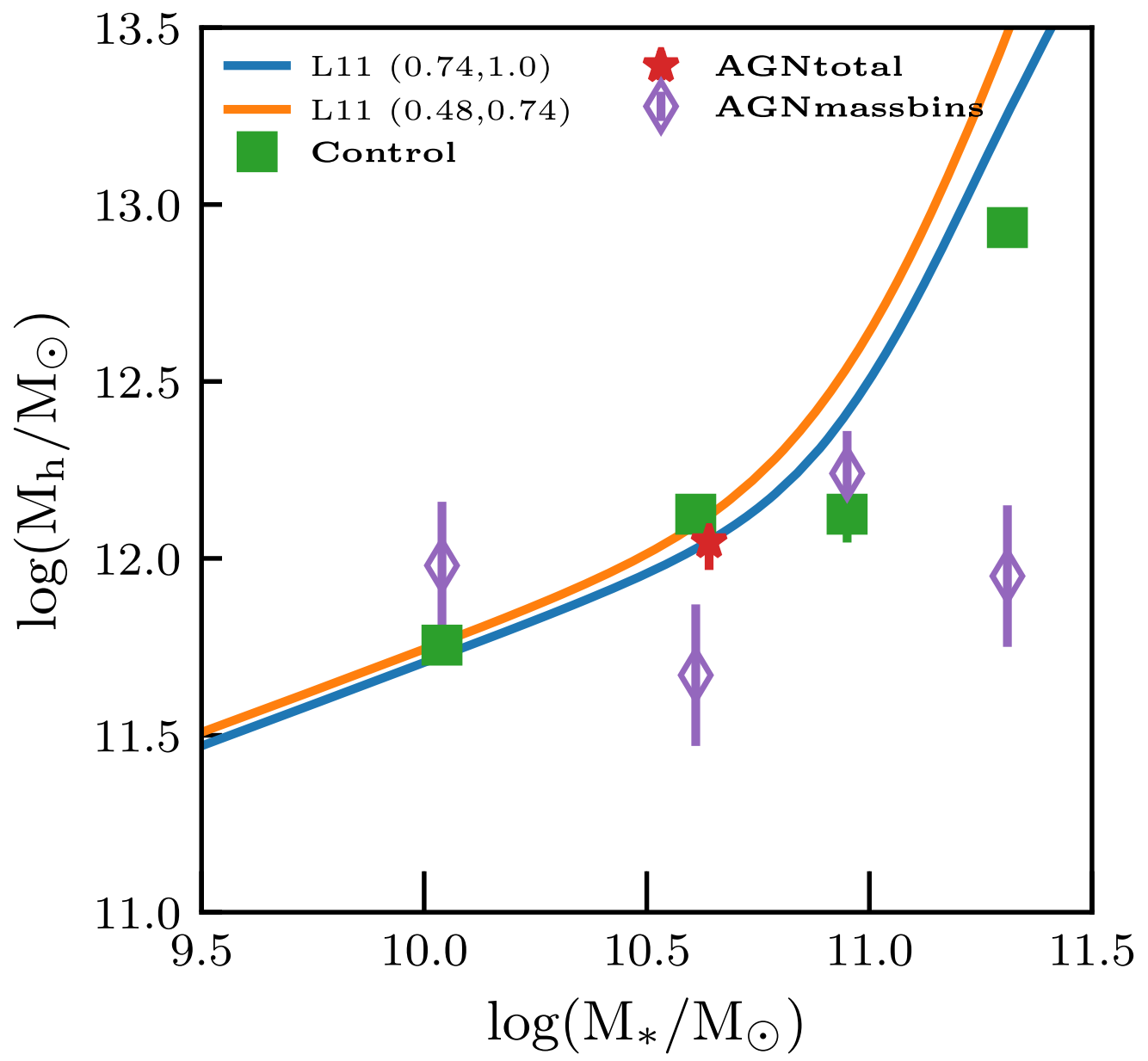

**Figure 8.** SMHR of the AGN and control galaxy samples. The theoretical lines are from Leauthaud et al. (2010; orange and blue solid lines). The green square data points show the control galaxy sample in four mass bins. The red star is the result based on the full AGN sample. The stellar mass bin AGN results are denoted as purple diamonds.

A picture may be emerging where SMBH accretion preferentially occurs in environments with plentiful gas supplies. This can be understood based on our comparison of luminous AGNs with a mass-matched control sample. For the higher stellar mass bins, the characteristic halo mass is significantly lower than the SHMR. And the satellite fraction $f_{\rm sat}$ for the AGN samples in our study shows no dependence on stellar mass. However, the constraint of the satellite fraction is not strong as shown in Figure 9. The satellite fraction of AGN samples lies between 10% and 35% for the different samples. Comparing to the weighted control sample, which shows a decreasing satellite fraction with increasing stellar mass, the satellite fraction has no strong dependence on stellar mass. The 16% satellite fraction agrees with the value from V. Allevato et al. (2019), who claim that their AGNs selected from the Chandra Cosmos Legacy Survey can reach up to 15% with halo mass around $\log(M_h/h^{-1}M_\odot) \approx 13.0$.

### 6.1. Consideration of Uncertainties on the Stellar Mass Measurements for Active Galactic Nucleus Host Galaxies

To address the issue of uncertainties on the stellar mass measurements of AGN host galaxies, we compare the values used here, based on SED fitting (A. D. Goulding et al. 2018), with those obtained independently through 2D decomposition of HSC images (J. Li et al. 2021). The latter are based on a sample of Type 1 quasars that do not represent the more obscured population which are included in this study. However, they have been tested to be robust through extensive simulations and represent the most likely problematic cases where a luminous quasar may impact the SED approach. Even though, J. Li et al. (2021) claim a 0.3 dex scatter as calibrated using simulations. Thus, we may find considerable scatter by comparing the two stellar mass estimations.

**Table 2**
Posterior of Parameters We Fitted to the Excess Surface Density Signals with Five Parameters

| Sample | $\log(M_h/h^{-1}M_\odot)$ | Concentration ($c$) | $f_{\rm sat}$ |
|---|---|---|---|
| AGN: all | $12.06^{+0.11}_{-0.12}$ | $4.86^{+1.9}_{-1.3}$ | $0.16^{+0.04}_{-0.04}$ |
| AGN: I | $11.85^{+0.28}_{-0.30}$ | $6.9^{+6.3}_{-3.7}$ | $0.20^{+0.08}_{-0.07}$ |
| AGN: II | $12.02^{+0.13}_{-0.13}$ | $5.8^{+2.5}_{-1.7}$ | $0.19^{+0.04}_{-0.04}$ |
| AGN: $0.2 \leqslant z < 0.65$ | $12.03^{+0.16}_{-0.17}$ | $6.0^{+3.4}_{-2.2}$ | $0.19^{+0.05}_{-0.05}$ |
| AGN: $0.65 \leqslant z < 0.80$ | $12.03^{+0.15}_{-0.17}$ | $7.7^{+4.8}_{-2.9}$ | $0.15^{+0.06}_{-0.05}$ |
| AGN: $0.80 \leqslant z$ | $11.46^{+0.35}_{-0.48}$ | $10.0^{+6.5}_{-6.1}$ | $0.27^{+0.12}_{-0.10}$ |
| AGN: $\log(M_*) \leqslant 10.4$ | $11.98^{+0.18}_{-0.24}$ | $4.18^{+3.5}_{-1.7}$ | $0.13^{+0.07}_{-0.06}$ |
| AGN: $10.4 < \log(M_*) \leqslant 10.8$ | $11.67^{+0.21}_{-0.19}$ | $12.15^{+5.4}_{-5.7}$ | $0.35^{+0.06}_{-0.07}$ |
| AGN: $10.8 < \log(M_*) \leqslant 11.1$ | $12.24^{+0.12}_{-0.19}$ | $4.7^{+2.9}_{-1.5}$ | $0.10^{+0.08}_{-0.06}$ |
| AGN: $11.1 < \log(M_*)$ | $11.95^{+0.20}_{-0.19}$ | $10.7^{+5.7}_{-4.7}$ | $0.22^{+0.08}_{-0.08}$ |
| AGN: low $L$ | $12.13^{+0.13}_{-0.15}$ | $5.1^{+2.5}_{-1.6}$ | $0.11^{+0.06}_{-0.05}$ |
| AGN: high $L$ | $11.87^{+0.28}_{-0.35}$ | $6.7^{+4.3}_{-2.5}$ | $0.21^{+0.05}_{-0.05}$ |
| Weighted galaxy bin 1 | $11.60^{+0.04}_{-0.04}$ | $12.7^{+2.7}_{-2.0}$ | $0.22^{+0.01}_{-0.01}$ |
| Weighted galaxy bin 2 | $11.97^{+0.05}_{-0.05}$ | $12.4^{+2.8}_{-2.2}$ | $0.21^{+0.02}_{-0.02}$ |
| Weighted galaxy bin 3 | $12.25^{+0.08}_{-0.08}$ | $10.50^{+3.4}_{-2.6}$ | $0.19^{+0.04}_{-0.03}$ |
| Weighted galaxy bin 4 | $12.77^{+0.05}_{-0.04}$ | $3.6^{+1.0}_{-0.6}$ | $0.05^{+0.05}_{-0.04}$ |

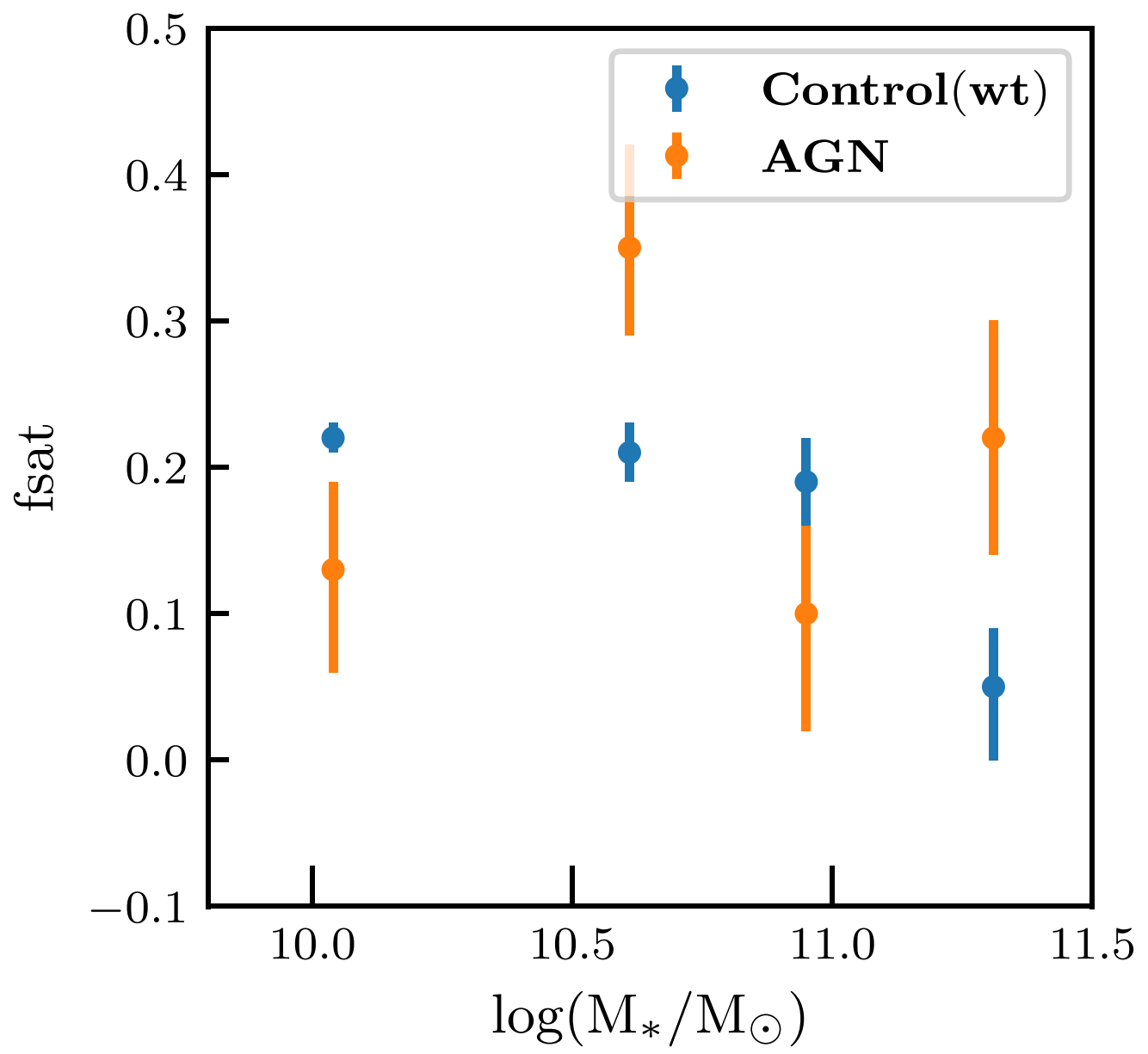

**Figure 9.** The satellite fraction of the control galaxy sample (blue dots) and the AGNs (orange dots) as a function of stellar mass. The AGN samples contain low/high stellar mass samples, the AGN sample as a whole, as well as the low/high stellar mass sample with larger separations, as in Figure 8.

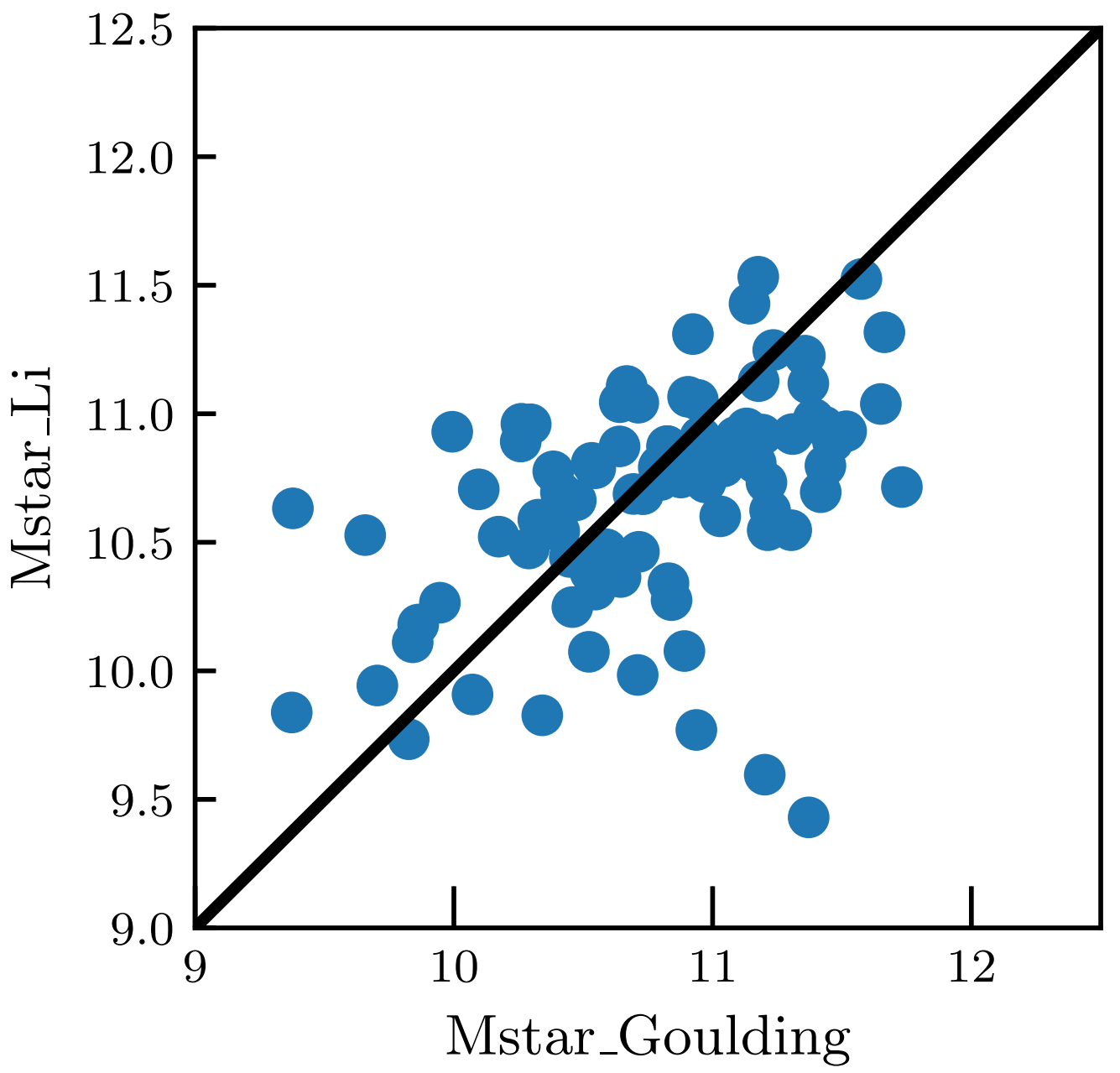

**Figure 10.** Comparison of stellar mass between weak lensing modeling and SED fitting. The $x$-axis shows the stellar masses of galaxies and AGNs and the $y$-axis gives the weak lensing modeled stellar mass.

In Figure 10, we show a comparison which demonstrates that this subsample follows the one-to-one relation thus there is no sign of a severe systematic offset. However, there is considerable scatter, 0.48 dex above $10^{10}$ $M_\odot$, which can impact our results if there is no further scatter between SHMR. However, our four stellar mass bins have much larger stellar mass differences than the 0.48 dex scatter. The largest stellar mass bin is about 1.3 dex larger than the lowest stellar mass bin, still their halo masses agree with each other at the $1\sigma$ level.

We carry out a Monte Carlo simulation to test the effect of the scatter of stellar mass on halo mass estimation. The Monte Carlo simulation simply takes the stellar mass function from O. Ilbert et al. (2013) and the SHMR from A. Leauthaud et al. (2012) to assign each Monte Carlo randomly sampled galaxy with a halo mass. We specifically select two massive stellar mass bins ($\log M_s \in [10.4, 10.80]$ and [10.80, 11.2] with a mean stellar mass of 10.61 and 10.97, respectively) and assign a 0.35 dex Gaussian random error to the two stellar mass bin mock catalogs and calculate the mean stellar mass and mean halo mass of each bin. We found that the 0.35 dex scatter only causes a 0.04 dex suppression of the mean halo mass. So our results are robust with different stellar mass estimations. This agrees with the recent study of X. Yang et al. (2021), who show that the scatter on the total luminosity of galaxy groups has little impact on the halo mass estimation.

The halo mass of the whole AGNs sample locates at the knee of the SHMR curve. After we separate the sample into high and low stellar mass bins, the halo mass of the high stellar mass bin deviates from the relation and is even lower than that of the control sample as shown in the upper panel of Figure 8.

In future, as more data become available from HSC-SSP, the SNR of the high-redshift AGN samples can be improved by a factor of 4 due to a larger survey region. We will continue this study with a larger data sample, which enables finer stellar mass bins.

## 7. Summary and Conclusions

We use an AGN sample from matched HSC and WISE catalogs (A. D. Goulding et al. 2018) to measure the summed galaxy–galaxy lensing signal in terms of the ESD. The ESDs of the AGNs are measured based on the HSC-SSP S16A shape catalog. The sample size of AGNs allows us to subdivide the sample into Types I and II, four stellar mass, three redshift bins, and high and low AGN luminosities. For comparison, we build a control sample of galaxies from the HSC-SSP S16A field matching the stellar mass and redshift distributions of the AGN sample based on a weighting scheme. The ESD of the control galaxies is also measured in four bins of stellar mass. The ESDs are summarized as follows.

1. We measure the galaxy–galaxy lensing signals, and ESDs around AGN samples and control galaxy samples. We found that the ESDs of AGNs as a whole agree with the second stellar mass bin control sample, indicating similar halo masses between the two samples.
2. We split the AGNs into four stellar mass and found the halo masses do not change as the halo masses in the control galaxy sample. Regardless of the 1.2 dex difference between the most massive AGN sample and the least massive ones, the halo mass are both around log $(M/h^{-1}M_\odot) = 12.0$.
3. We also try different sample selections based on luminosity, type, and redshift. We did not find any significant differences in these subsamples.

We further build a model to extract the information of halo mass, concentration, satellite HOD, and satellite fraction. We run MCMC using `emcee` to sample the posterior distribution of each parameter. We find that the halo mass is very sensitive to the ESD as expected while the concentration parameter and amplitude in the simple power-law satellite HOD are not. The satellite fraction is relatively well constrained in the control sample thanks to the high SNR ESD measurements. We summarize our main results as follows.

1. The mean halo mass of the full AGN sample is $\log(M_h/h^{-1}M_\odot) = 12.06^{+0.11}_{-0.12}$. The AGNs are located at the knee of the SHMR, where star formation is the most efficient.
2. The mean halo mass of the galaxy control sample follows the trend of the calibrated SHMR as in Figure 8. This shows our model reasonably extracts the halo mass information. One interesting discovery in our work is that the mean halo mass of the highest stellar mass AGN sample $\log(M_h/h^{-1}M_\odot) = 11.95^{+0.20}_{-0.19}$ is about $5\sigma$ lower than the calibrated SHMR. Meanwhile the mean halo mass $\log(M_h/h^{-1}M_\odot) = 11.98^{+0.18}_{-0.24}$ of the low stellar mass AGN sample is comparable to the established SHMR.
3. The Type I and Type II AGNs have marginally distinguishable halo masses given the same stellar mass and redshift distributions. The mean halo mass of Type I AGNs is $1\sigma$ less than that for Type II AGNs. The mean halo mass of high AGN luminosity sample is also lower than for the low AGN luminosity sample, but due to the low SNR the two halo masses are still consistent to within the $1\sigma$ error. We also reweight the galaxy control sample to the $M_*-z$ distribution of Type II AGNs and the low-luminosity AGN sample. We do not find significant difference between Type II AGNs, low-luminosity AGNs, and the galaxy control sample.
4. The satellite fraction of the control sample shows an decreasing trend as a function of stellar mass. The higher the stellar mass, the smaller the satellite fraction, which is not the case for the AGN samples. The satellite fraction of AGN samples with different stellar mass bins does not show a similar trend as the control sample. The value is ranging from 10% to 35%, indicating that AGNs are not sensitive to their environment.

In the future, we expect to measure higher SNR galaxy–galaxy lensing signals from HSC-SSP Y3 shape catalog. With the higher SNR signals, we are able to explore more the halo properties of AGNs with a finer binning scheme. Besides the optical–IR-selected AGNs, we can also compare the X-ray-selected AGNs with the sample we used in this work.

## Acknowledgments

We thank the anonymous referee for the valuable feedback that significantly improved the content of our work. We also thank Zheng Zheng from the University of Utah for valuable discussions of HOD modeling and Qinxun Li from the University of Science and Technology of China for the HOD code sanity test. W.L. acknowledges the support from NSFC (NO.12192224), the National Key R&D Program of China (2023YFA1608100, 2023YFA1608100 and 2021YFC2203100), the 111 Project for "Observational and Theoretical Research on Dark Matter and Dark Energy" (B23042), the Fundamental Research Funds for the Central Universities (WK3440000006), and World Premier International Research Center Initiative (WPI Initiative), MEXT, Japan. J.S. is supported by JSPS KAKENHI grant No. (JP18H01251 and JP22H01262), and the World Premier International Research Center Initiative (WPI Initiative), MEXT, Japan. All numerical computations were performed on the GFARM computing cluster at Kavli IPMU, the University of Tokyo.

## ORCID iDs

Wentao Luo https://orcid.org/0000-0003-1297-6142
John D. Silverman https://orcid.org/0000-0002-0000-6977
Surhud More https://orcid.org/0000-0002-2986-2371
Andy Goulding https://orcid.org/0000-0003-4700-663X
Takahiro Nishimichi https://orcid.org/0000-0002-9664-0760
Lalitwadee Kawinwanichakij https://orcid.org/0000-0003-4032-2445
Junyao Li https://orcid.org/0000-0002-1605-915X
Xiangchong Li https://orcid.org/0000-0003-2880-5102
Elinor Medezinski https://orcid.org/0000-0001-7007-2358
Masamune Oguri https://orcid.org/0000-0003-3484-399X